# Title: Graph-based variant discovery reveals novel dynamics in the human microbiome

**Short Title:** Variant detection from assembly graphs
**One Sentence Summary:** Linking assembly graph motifs to biological variants, like strain-level differences, that affect microbial function

**Authors:** Harihara Subrahmaniam Muralidharan[1,2], Jacquelyn S Michaelis[2], Jay Ghurye[1,2], Todd Treangen[3], Sergey Koren[4], Marcus Fedarko[5], and Mihai Pop[1,2*]

**Affiliations:**

[1]Department of Computer Science, University of Maryland, College Park, Maryland, USA

[2]Center for Bioinformatics and Computational Biology, University of Maryland, College Park, Maryland, USA

[3]Department of Computer Science, Rice University, Houston, TX, USA

[4]Genome Informatics Section, Computational and Statistical Genomics Branch, National Human Genome Research Institute, National Institutes of Health, Bethesda, Maryland, USA

[5]Department of Computer Science, University of California, San Diego, California, USA

*Correspondence to: mpop@umd.edu

## Abstract

Sequence differences between the strains of bacteria comprising host-associated and environmental microbiota may play a role in community assembly and influence the resilience of microbial communities to disturbances. Tools for characterizing strain-level variation within microbial communities, however, are limited in scope, focusing on just single nucleotide polymorphisms, or relying on reference-based analyses that miss complex functional and structural variants. Here, we demonstrate the power of assembly graph analysis to detect and characterize structural variants in almost 1,000 metagenomes generated as part of the Human Microbiome Project. We identify over nine million variants comprising insertion/deletion events, repeat copy-number changes, and mobile elements such as plasmids. We highlight some of the potential functional roles of these genomic changes. Our analysis revealed striking differences in the rate of variation across body sites, highlighting niche-specific mechanisms of bacterial adaptation. The structural variants we detect also include potentially novel prophage integration events, highlighting the potential use of graph-based analyses for phage discovery.

## Introduction

Genomic changes that occur within a short evolutionary timeframe, even within a few generations, play an important role in the adaptation of bacteria to their environment, particularly in the context of host-microbe and microbe-microbe interactions [1]. While such genomic changes have been studied in a relatively small set of organisms, little is known about the extent and rate at which genomic changes occur within natural microbial communities, and how genomic variants correlate with the environmental context within which the microbial communities exist. Some evidence also exists that certain structural genomic variants within human-associated microbiota may be associated with human disease risk factors, including body weight, cholesterol, and other markers of metabolic disease [2]. Structural variants detected in the human gut microbiome have been further correlated with altered metabolism of bile acids, are involved in different signaling pathways, and may play a role in inflammatory bowel disease [3, 4].

The detection of genomic variants between different organisms in a microbial community is non-trivial. Many current approaches are reference-based, mapping reads against existing genes or genomes to identify single-nucleotide polymorphisms (SNPs) and then using multi-locus sequence typing (MLST) schemes [5, 6] or correlations in the frequency of variants across samples [7, 8] to characterize strains and their abundance. Reference-independent approaches aimed at characterizing uncultivated microbes rely on detecting changes in core and accessory genes within contig bins [9, 10]. Although important for differentiating taxa, these strain level variants defined by SNPs miss structural information, like gene loss, gain, or transfer, that underlie important host phenotypes.

Only two approaches have specifically targeted the characterization of structural variants in microbiome data [2, 11]. Both approaches rely on reference genome sequences and infer structural variants based on the alignment of metagenomic reads to these references. In [11], the focus is on genomic regions that can invert (invertons), leading to changes in gene expression, e.g., when the inverted region contains a promoter. In [2], the focus is on insertion/deletion events with respect to the reference sequence, inferred from the depth of coverage of read mappings against the reference.

To expand the scope of structural variants that can be characterized and extend analyses to organisms for which no reference sequence is available, we propose here to use the information contained in genome assembly graphs. Graphs are commonly used to assemble overlapping reads into contigs, and to orient and link together contigs into scaffolds using paired-end read information. While many assemblers discard ambiguity within the graph when generating contigs, variation-aware tools that preserve graph structures [12-15] can reveal patterns associated with complex functional and structural changes in microbial communities. Here we use the variant-aware, *de-novo* scaffolder MetaCarvel [12] to catalogue six different types of structural variants in data generated by the Human Microbiome Project (HMP) [16]. We also describe a normalization approach that allows the accurate comparison of structural variants across samples with different depth of coverage, revealing differences in the genomic variation present in the microbiota associated with different body sites.

# Results

## Structural variants are highly prevalent and vary across micro-environments

We focus on six types of structural variants that yield recognizable patterns in assembly graphs (Figure 1A). Insertion/deletion (indel) events and localized regions of variation appear as "bubbles" in the graph comprising two parallel paths. Tandem repeats lead to simple cycles located in the middle of a path in the graph. Intra-genomic and inter-genomic repeats yield nodes in the graph that are flanked by multiple different contigs. We also identify putative plasmids (graph components that comprise a single cycle) as well as complex variants represented as "bubbles" with multiple parallel paths.

Our analysis revealed over nine million structural variants in 934 HMP samples from six different body sites **(Data S1).** The extent of DNA contained within variants had a median length of 1.2 kbp per variant (min 0.5 kbp, mean 2.5 kbp, max 546.3 kbp) (Figure S1).

Because sample sequencing depth can impact the number of structural variants that can be detected in the graph, we developed ANCHOR (v**A**riant **N**ormalization by **C**overage and dept**H O**f **R**eads, see Methods) to normalize the graph structure across samples with varied depth of coverage. ANCHOR reduces bias by using statistical methods to determine the mean number of variants that would be detected if reads from one sample were down-sampled to the average sequencing depth of another sample. Briefly, ANCHOR determines the probability that each edge in a bubble, and ultimately the bubble itself, would survive at a lower depth of coverage. The algorithm runs multiple iterations, using probabilities to determine the presence of each edge in a graph and to count the number of bubbles that are retained. The survival characteristics of a bubble are not obvious **(**Figure 1B**)**. For instance, after down-sampling, an indel can remain an indel, disappear, or transform into a simple or complex structural variant, depending on how the structure of the graph changes as coverage is reduced.

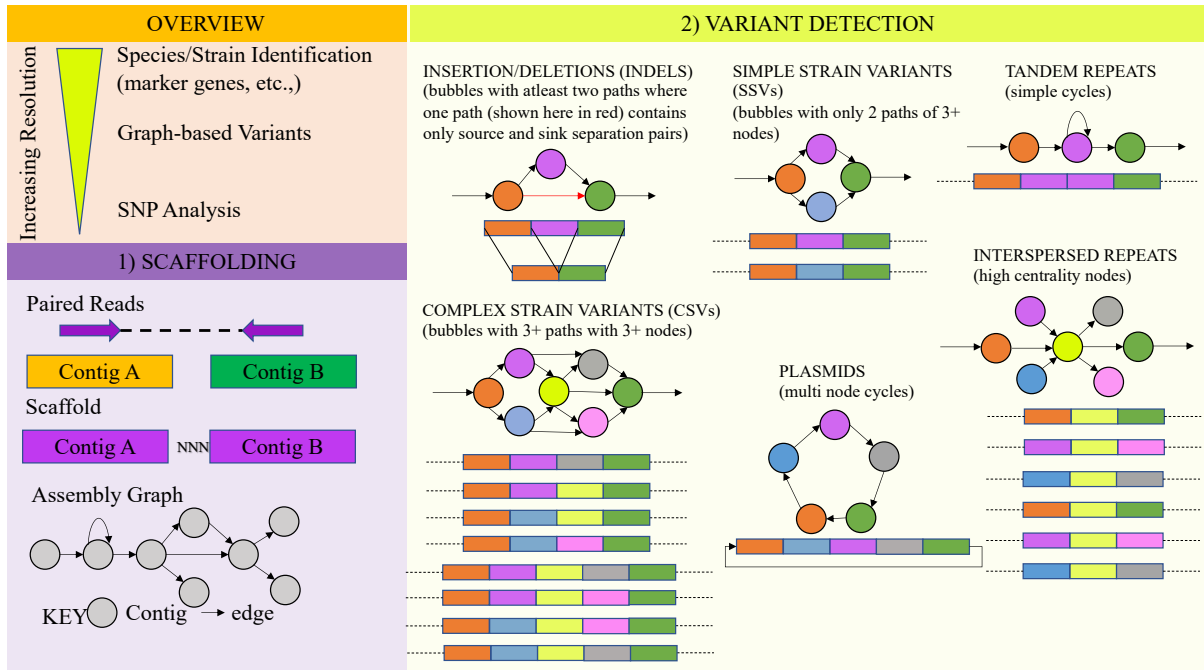

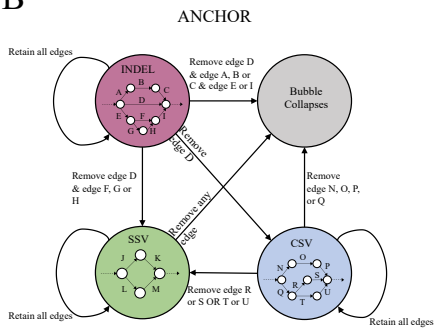
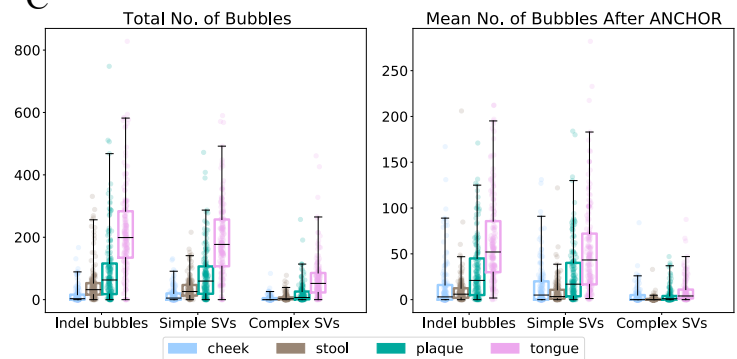

*Figure 1. Overview of structural variant detection and analysis. (A) Structural variants are detected from scaffolding assembly graphs where contigs are nodes and edges denote sufficient mate pair support between contigs. Genomic segments are highlighted in different colors which also highlight the corresponding graph nodes. (B) The ANCHOR algorithm was used to normalize and compare variants at different body sites from samples of differing sequencing depths. Each circle in this plot represents a bubble variant type example and potential outcomes when down sampling. (C) Box plots highlight the total number of detected bubble variants within HMP oral and gut samples (left) and the mean number of bubble variants after down sampling to the mean number of reads within the cheek samples (right).*

We used ANCHOR to normalize the structural variant content of samples from stool, tongue, buccal mucosa, and supra-gingival plaque. As a basis for normalization, we used the buccal mucosa samples which had the lowest overall sequencing depth. Even after normalization, oral samples had greater numbers of indels and other strain variants than stool samples **(Figure 1C)**. This result differs from previous OTU-based alpha diversity analyses [16] but is consistent with recent strain-level analyses [17] and deeper analyses of diversity across body sites [18]. We hypothesize that the increased diversity observed in the oral cavity may be due to nutrient availability; the gut, especially stool, which is a composite of digested food moving throughout the intestinal tract, contains more nutrients than the mouth. When nutrients are sparse, microbes may need to develop different strategies to survive, leading to a higher diversity even within closely related strains.

# The functional potential contained in structural variants differs across body sites

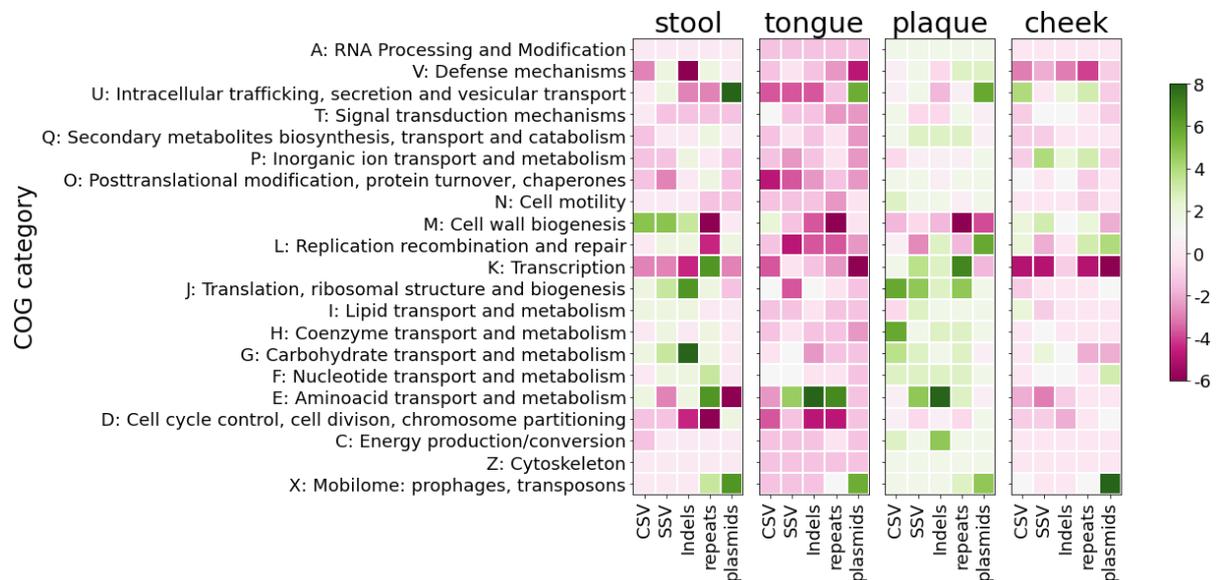

*Figure 2. The COG category of differentially abundant genes broken down by structural variant type and body site. We identified the top and bottom 25 genes with respect to their enrichment/depletion within a particular variant type with respect to the entirety of the body site, and recorded the number of times a particular category was observed in the top or bottom list. When a COG category was found in both the top and bottom groups of genes, the corresponding numbers were subtracted to yield the information shown in the figure.*

We used eggnog mapper [19] to functionally annotate the contigs found within structural variants, and focused our analysis on those genes that appear to be over- or under-represented within structural variants in different body sites. The enrichment or depletion of genes within particular types of variants was defined in terms of the difference between the frequency of a functional label within a type of variant and the frequency of the label within the entire body site (see Methods).

Plasmids were enriched in genes involved in replication, chromosome partitioning, as well as toxin/anti-toxin systems, all functions that are important for replication and persistence. Surprisingly, we also identified in plasmids an enrichment in phage-related functions. There has been recent interest in a class of temperate phages that mimic plasmids during their lysogenic phase [20-22] and we sought to determine whether such elements may be present in the HMP data. We downloaded the phage-plasmids identified in [21], and aligned them to the plasmids we identified in HMP stool samples. We found full-length matches between 5 plasmids and previously characterized phage-plasmids, as well as a number of shorter matches, supporting the idea that the HMP data contain known phage-plasmids.

Interspersed repeats (as defined by our analysis) represent regions that occur within multiple genomic contexts, including either intra-genomic repeats or mobile genetic elements that are inserted in different locations in the genomes found in a sample. This variant type was enriched in functions associated with the mobilome (Figure 2) in all body sites. Biological functions that are essential for the survival of cells, such as cell cycle control and cell wall biogenesis were depleted from interspersed repeat regions, consistent with the expectation that such functions are less likely to be laterally transferred. Similarly, insertion/deletion (indel) regions were depleted of essential functions.

Some noted differences between body sites included a stronger association with variants (particularly indels and simple structural variants) of functions associated with cell wall biogenesis in stool samples, while the same functions were depleted from the variants in plaque and cheek samples. Functions that are depleted from variant regions may be critical for survival in a particular environment. Thus, this observation suggests that cell wall related functions are important for survival in the oral cavity, perhaps due to the fact that many microbes exist within biofilms in this body site. Conversely, functions that are associated with variants my indicate that organism may need to innovate in order to remain competitive within the environment. Changes in cell wall related functions may allow microbes to survive in stool, perhaps as a mechanism to evade phages, the immune system, or antimicrobials produced by other microbes. This hypothesis is further supported by the observation that functions related to defense mechanisms are strongly depleted from indels in stool (in contrast to other body sites), suggesting such functions are important in this body site.

## Variation-aware bins show greater genus-level diversity in the oral cavity compared to the gut

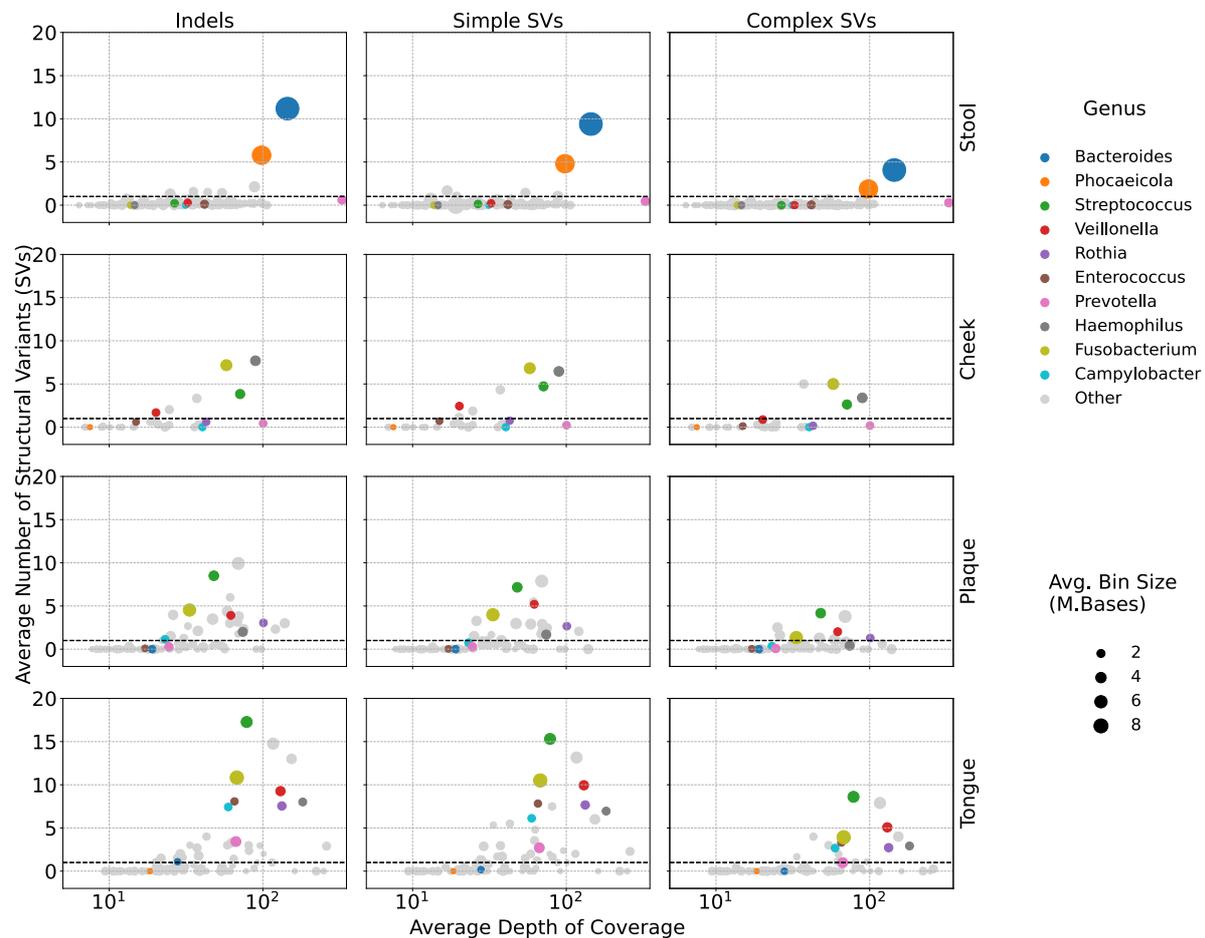

*Figure 3. Relationship between depth of coverage and number of structural variants detected across body sites and taxonomic groups after normalization with ANCHOR. Each dot represents the average size of the bins(size) identified that belong to a particular bacterial genus (color), faceted by microenvironment and SV type. The x-axis denotes the average depth of coverage of bins in base pairs per genera, while the y-axis shows the average number of bubbles per bin.*

To further explore structural variant diversity of the different oral body sites and stool samples, we used binnacle [23] to calculate the coverage of the graph scaffolds and metabat2 [24] to bin them. Further we used ATLAS [25] to assign taxonomy to the bins identified from the oral and stool samples and focused on genera found in at least five bins (Figure 3). As noted previously, even after matching the differences in structural variant counts with ANCHOR, we observed that bins in the oral cavity contained more structural variants than stool samples. Oral cavity bins with multiple structural variants belonged to *Haemophilus*, *Fusobacterium*, *Streptococcus* and *Veillonella* spp., while stool bins with multiple structural variants belonged primarily to *Bacteroides* and *Phoecaeicola* spp. While in stool samples, the number of variants was largely correlated with depth of coverage, this correlation was substantially less marked in oral samples (Figure 3).

Despite comparable sequencing depths of the *Streptococcus* bins in the oral samples, we observed that the number of structural variants in the bins varied across body sites even when restricted to this single genus. To further understand the differences in the genetic composition of the *Streptococcus* bins, we predicted proteins within the bins and assigned functions using EggNOG mapper. We identified genes that were significantly enriched in the structural variants within each of the three sub-sites of the oral cavity (Fisher's exact test, FDR < 4e-06). These genes were broadly associated with functions such as translation, uptake of metabolites, nucleotide biosynthesis, carbohydrate utilization, DNA damage repair, cell wall biosynthesis, and DNA replication. We also noted an enrichment of the CcpA gene, responsible for catabolite repression.

## Structural variants can reveal novel phages

Our initial analysis of the functional elements enriched within the interspersed repeats identified by MetaCarvel suggested that they contain prophage or phage-like elements. To more accurately evaluate the phage content of intersperse repeats, we relied on VirSorter [26] to identify contigs that likely belonged to phages or prophages. As expected, interspersed repeats were significantly enriched in known phages (Fisher exact test-p-value < 0.005) compared to the other structural variants. As a specific example, we sought to determine whether we could identify the genomic contigs associated with *CrAssphage*, a class of phages that infect *Bacteroides* organisms, and that are ubiquitously found in human gut samples and which were initially discovered through metagenomic sequencing [27]. We aligned the interspersed repeat contigs against the *CrAssphage* genome using Minimap2 [28] and retained only high-quality alignments (nucleotide identity > 75% and alignment length > 1kbp). More than 99% of the *CrAssphage* genome could be recovered by using only interspersed repeat contigs.

To further explore the ability of assembly graph analysis to discover novel phage and phage-like genomic segments, we developed a pipeline PIRATE (Phage Insertion fRom Assembly-graph varianT Elements, described in methods). We aligned all the interspersed repeat contigs against the NCBI *nt* database and identified long contigs (> 3kbp) that did not align well to database sequences (breadth of alignment coverage < 50%). This resulted in the identification of 105,009 candidate contigs out of 541,731 contigs. To identify sequences that appear in multiple samples (and therefore are less likely to represent experimental artifacts), we clustered the contigs into 9,793 clusters. We, then, assigned the longest contig in each cluster as the representative, and analyzed the corresponding sequence by predicting genes and assigning functions using the EggNOG mapper [19]. Over a quarter of the clusters (2,546 out of 9,793) had at least one hit to a phage-related protein domain.

The clusters identified by PIRATE ranged in prevalence from one to 78 samples (Figure 4A). We highlight a cluster (Figure 4B) comprising 84 contigs from 64 samples, along with the functional annotations of key genes suggesting the phage origin of this cluster.

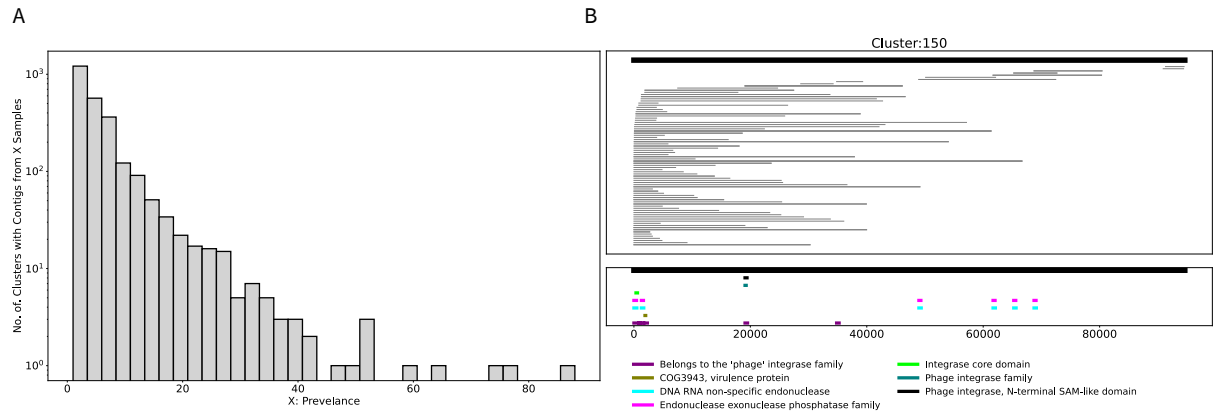

*Figure 4 Phages identified in interspersed repeats. (A) Histogram showing the number of samples within which clusters are found. (B)The protein domains predicted on the contigs from cluster 150 identified by PIRATE. The black bar at the top represents the cluster representative of cluster 46 and the grey bars shows the tiling of 127 contigs from 41 stool samples.*

# Discussion

Prior studies of strain-level variation with microbial communities have relied on either marker gene based or SNP-based variant discovery which are all dependent on references and have limited ability in discovering novel genomic information. In this work, we provide insights into of genomic variation within microbial communities in a data-dependent and reference free manner. We have shown that assembly graph analyses can enable the exploration of a broader set of structural variants than possible through reference-based analyses within host-associated microbiota, and provided an initial analysis of over 9 million structural variants discovered within the samples generated by the Human Microbiome Project.

Our analysis adds to ongoing efforts to characterize and interpret the diversity of host-associated microbial communities (e.g., [18, 29, 30]), an important consideration given attempts to associate microbiota diversity with host health [29]. Such analyses, however, are strongly confounded by the depth of sequencing of each sample since the ability to characterize structural variants depends on the contiguity of the assembly. For host-associated microbiome samples, the depth of sequencing is tightly connected with the amount of host DNA found in a sample, thus analyses comparing samples from different body sites require careful normalization. Here we have proposed a down-sampling based approach for normalizing structural variant information, allowing for a more accurate comparison across samples.

This analysis revealed a marked difference in the number of structural variants detected in oral samples, compared to stool samples, with a more significant extent of variation evident in oral samples. Looking specifically at a single taxonomic group, oral members of the *Streptococcus* genus harbored more structural variants than those found in the gut despite occurring at approximately the same depth of coverage in both types of samples. A possible explanation is that survival in the relatively nutrient-poor oral environment requires constant adaptation and functional innovation, while a "winner takes all" strategy is more successful in the nutrient-rich gut environment.

We have also shown that analysis of structural variants, particularly interspersed repeats, can reveal novel mobile elements. Interspersed repeats are genomic regions that appear in different genomic contexts within the community, consistent with a mobile element or prophage region. Many of such regions we identified had limited or no sequence-level similarity with publicly available data-sets, however a significant fraction of these regions had distant hits to phage proteins, suggesting they may represent as of yet uncharacterized phage families. This finding further demonstrates the untapped potential of metagenomic sequencing for discovering new genomic elements, and the graph-based approach embodied in PIRATE complements the cross-assembly technique used to previously discover the *CrAssphage* family. Our analysis has also confirmed the presence of phage-plasmids among the plasmid sequences we identified in the HMP data. The current knowledge of phage-plasmids is based on analyzing known phage and plasmid sequences, and our study suggests that the analysis of metagenomically-derived sequences may lead to further discoveries.

We hope that our manuscript has highlighted the promise of studying microbial communities through the lens of structural genomic variants which can reveal information about the functional adaptation within environmental niches as well as highlight previously uncharacterized mobile genetic elements. Substantially more research is, however, needed to further improve our ability to characterize structural variants within metagenomic assemblies and to compare the variant-ome across samples at different levels of coverage. Furthermore, developing techniques for longitudinal analysis and for integrating variant information with meta-transcriptome and metabolome data will be necessary to disentangle the functional impact of strain variation in microbial communities.

# Materials and Methods

## Scaffolding, Variant Detection, and Normalization

HMP assemblies and read sets were downloaded from NIH Human Microbiome Project (**Data S1**) and MetaCarvel v1.1 was used to scaffold and detect variants using default parameters. We used ANCHOR (see below) to normalize all stool and oral samples to the buccal mucosa samples, setting $N_H$ to the number of reads in each of the stool, tongue, and plaque samples and $N_L$ to the average number of buccal mucosa reads, and running for 1,000 iterations per sample.

We categorized the structural variants identified by MetaCarvel into 3 major types namely indels, simple structural variants (SSV), and complex structural variants (CSV). We define indel variants to be bubbles with exactly 2 paths and such that one of the paths contains just 2 nodes; the source and sink of the bubble. Simple structural variants are bubbles with exactly 2 paths with at least 3 nodes between the source and sink. Complex bubbles are bubbles with multiple paths of length at least 3 between the source and sink nodes. We also identified circular sub-graphs which are consistent with the presence of plasmids in the samples.

## vAriant Normalization by Coverage and deptH Of Reads (ANCHOR)

Since sequencing depth can impact the number of structural variants detected, we propose a simple probabilistic model to count the number of bubbles after normalizing sequencing depths between two groups of samples. We perform the following procedure to account for the difference in sequencing depth without having to re-analyze down-sampled data sets:
1. Let the sequencing depth of a sample with higher depth be given by $N_H$ and that of lower depth be given by $N_L$. The probability of each mate pair edge existing between two nodes is given by $p = \frac{N_L}{N_H}$.
2. MetaCarvel links two contigs by an edge if there are at least $m_{min}$ mate pair edges. Assuming there are $m$ mate pair edges between two contigs the probability of an edge existing between two contigs can be computed using a simple Bernoulli model:

$$P = \sum_{i=m_{min}}^{i=m} \binom{m}{i} p^i (1-p)^{m-i}$$

3. For all bubbles in the sample, gather the edges that are present in the bubble.
4. Retain each edge with a probability $P$. We check what happens to the bubble after the edge deletion phase, i.e., the bubble, depending on its type might either transition into a simpler structure, disintegrate entirely or survive in its current state.
5. Repeat this process for all the samples and count the number of variants.
6. Repeat the above process over 1,000 of trials to determine the uncertainty in the estimated counts.

## Taxonomic and Functional Annotation

When characterizing variants, taxonomy was assigned to contigs within each variant using ATLAS [25]. Genes were predicted from contigs using Prodigal [31] in metagenome mode and eggnog mapper [19] was used to search for the functional roles of contigs within variants.

For a body site b, variant type v, and gene g, we compute the following quantities,

$$R_{v,b}^g = \frac{counts(g|v,b)}{counts(v,b)} \quad R_b^g = \frac{counts(g|b)}{counts(b)}$$

The $R_{v,b}^g$ reflects the ratio of the number hits to the gene g in body site b, and variant v to the total number of hits in the body site b and variant v across all genes. The $R_b^g$ reflects the ratio of the number hits to the gene g in body site b to the total number of hits in the body site b across all genes. We then computed the z-statistic of the two ratios. A high z-statistic signifies that the gene is enriched in the variant compared to the rest of the genome. We then identified the top 100 genes based on $counts(g|v,b)$ and selected top and bottom 25 genes based on the z-statistic for further analysis.

## Metagenomic Binning

Binnacle [23] is a variant aware binning method that estimates coverages on graph-based scaffolds identified by MetaCarvel [12]. We used the coverages estimated by binnacle and the sequence composition as features to identify metagenomic bins using MetaBAT 2 [24]. We assigned taxonomy to the bins using ATLAS [25]. We assigned functions to the contigs in the Streptococcus bins using EggNOG.

## Phage Detection with PIRATE

We gathered contigs from all gut metagenomic samples that were tagged as interspersed repeats by MetaCarvel. We aligned all the contigs against the NCBI non-redundant *nt* database using BLAST 2.13.0 and deemed any contigs with a breadth of coverage less than 50% as "novel" since they are not well represented by databases. Breadth of coverage was defined as the fraction of bases in the query sequence that are aligned to the database. We also filtered the results to retain only contigs greater than 3,000 bp. The workflow used to identify the candidate contigs is available in PIRATE Preprocess.py. To identify novel phage elements:

1. We performed an all-vs-all alignments of the filtered "novel" contigs using MiniMap2 [28].
2. We sort the contigs in decreasing order of their lengths and starting with the largest contig as cluster representative, we recruited to that cluster all contigs that aligned with an identity of at least 70% to the representative contig. Further, the cluster members also had to have a query coverage of at least 80% with respect to the representative contig.
3. We predicted genes on the contigs and assigned functions to the genes using EggNOG mapper.

# Funding

JG, JSM, HM, MF, and MP were partly supported by the NIH, award R01AI100947 to MP.

# Authors Contributions

MP conceived the research project. JG designed and implemented the algorithm, with the help of JSM, HM, and MP. JG, JSM, and HM analyzed the variants in the HMP dataset. HM created ANCHOR. SK, MF, and TT helped with data analysis. JG, JSM, HM, and MP wrote the manuscript. All authors read and approved the final manuscript.

# Competing Interests

The authors declare no competing interests.

# Data and materials availability

All data analyzed was obtained from the Human Microbiome Project, deposited to NCBI under BioProject PRJNA48479.

# Supplementary Materials

Data S1: Information about HMP samples analyzed.

Supplementary Figures

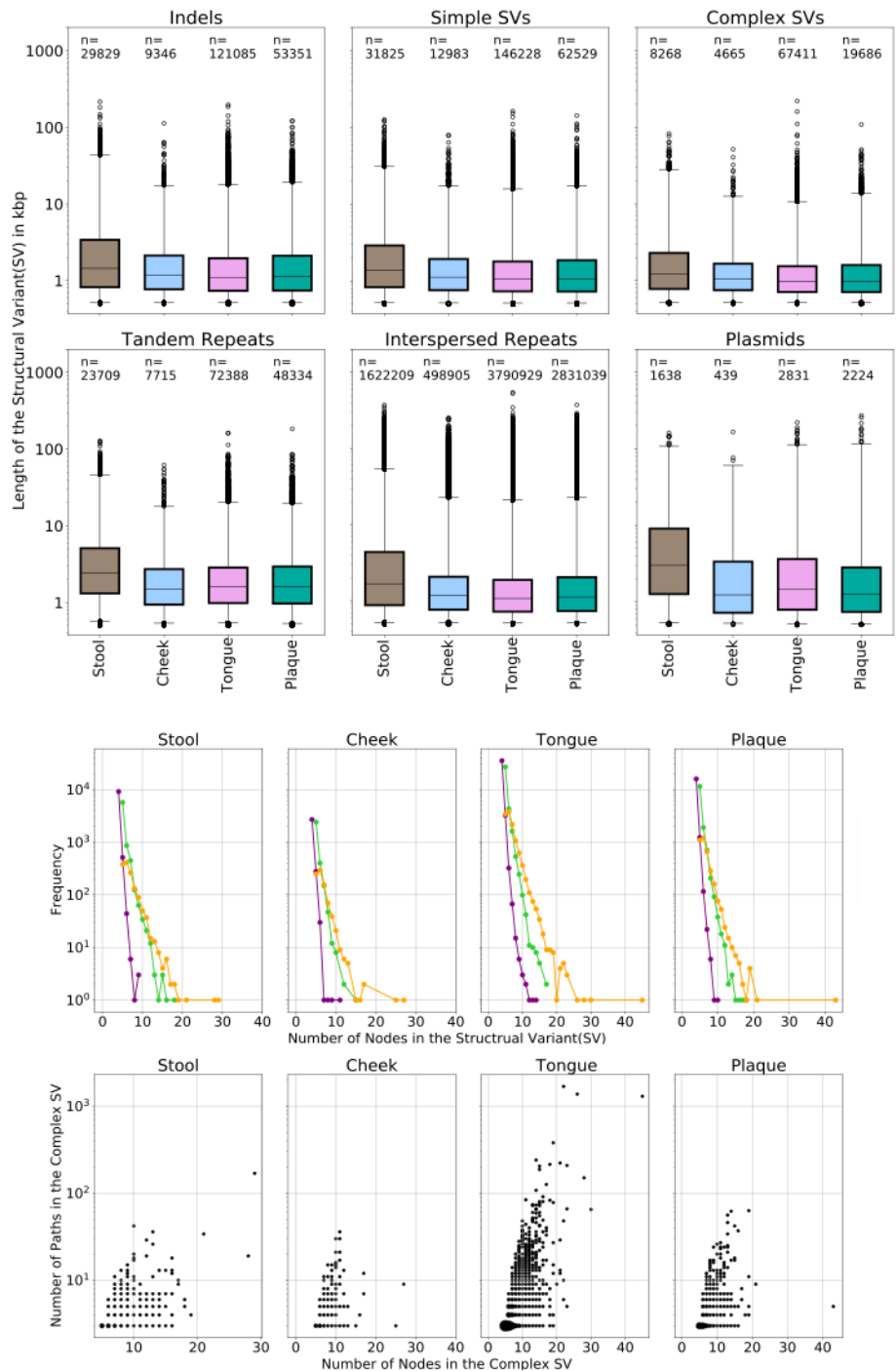

*Figure S1. (Top) Boxplots showing lengths of detected SVs (y-axis) by body site (x-axis) and variant type (group). Number of variants included in each plot is depicted at the top. (Middle) Total number of bubble SVs by number of nodes (contigs) in the bubble, separated by bubble type and body site. (Bottom) Total number of complex SV per body site by number of nodes in the bubble (x-axis) and number of paths in the bubble (y-axis).*